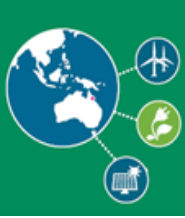
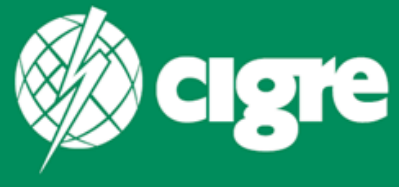

# Symposium Paper

## Paper information

## Summary

Communication-enabled devices in the grid make it more prone to cyberattacks. Attackers use different mechanisms, such as denial of service (DoS), false data injection (FDI), stealthy attacks, and replay attacks. They can use different types of attacks simultaneously, known as hybrid attacks, to cause more damage. For instance, they can use DoS and FDI attacks simultaneously on Volt-Var control resources of distribution systems such as on-load tap changers (OLTC), capacitor banks, and distributed energy resources (DER). Cyberattacks can affect the Volt-Var control and cause unbalanced voltage, overvoltage, and undervoltage. Only a few research works have studied hybrid cyberattacks, and hybrid cyberattacks on the Volt-Var control system have not been sufficiently studied. In this paper, a data-driven mitigation and detection method for a hybrid attack, DoS, and FDI attacks, on Volt-Var control is developed. DoS attack is detected if a user uses over 60% of the feeder remote terminal unit capacity. It is mitigated by replacing the lost value estimated by an artificial neural network (ANN) and disconnecting the attackers. The FDI is detected if the difference between the estimated value by ANN and the received measurement packet is more than 10% since the accuracy of the ANN is about 90.3%. If the difference between the estimated and the received value is greater than 10%, the estimated value is replaced with the received measurements. The ANN-based algorithm is implemented in the cyber layer of the control center. Linearized Volt-Var control is used in the distribution grid since it has thousands of nodes. This paper implemented a linearized Volt-Var control algorithm. The physical layer is implemented in PSCAD, the cyber layer is implemented using MATLAB Simulink, and the physical and the cyber layer are linked using the PSCAD cosimulation environment and Matrikon OPC for linking the cyber and physical layers. The proposed method successfully detects and mitigates two cases of hybrid cyberattacks as tested using the IEEE 13-bus test feeder system.

## Keywords

Cyberattack, detection, hybrid attack, mitigation, Volt-Var.

## Introduction

By developing smart grid and using more communication-enabled devices, cyberattacks have become a major concern since attackers can use these devices to cause power outages and disturbances. Cyberattacks can be applied to different parts of the grid, such as voltage control devices that are used to maintain the voltage within the limit and cause voltage disturbances. Examples of voltage control devices in distribution are onload tap changers (OLTC), voltage regulators, DERs, and capacitor banks, which are used to maintain the voltage within limits [1]. Moreover, attackers use different types of cyberattacks, and more advanced types of attacks have been introduced that need to be studied.

Cyberattacks on the Volt-Var control can cause voltage disturbances in the distribution system by changing the reactive power injection and OLTC tap positions. Reference [2] shows that discoordination in Volt-Var control and injecting a higher amount of reactive power can cause overvoltage. Reference [3] formulates a simple detection and mitigation method based on central voltage measurement for FDI attacks that can cause voltage violations on the distribution grid having PV. However, it only works well if a small number of sensors are under attack, and its efficiency decreases when more sensors are under FDI attack. A simple detection method is proposed that works well only if a small number of sensors are under attack.

Attackers use different types of cyberattacks on the electrical grid, such as FDI, stealthy, DOS, ARP spoofing, and man-in-the-middle attacks. Additionally, attackers have been using more advanced cyberattacks in recent years. Some articles have developed novel attack methods to challenge the intrusion detection system and enhance the cybersecurity of systems. For instance, Reference [4] introduces a novel stealthy attack strategy that passes through the Volt-Var optimization system without being detected by formulating it as a one-level optimization problem, where the voltage measurement for all nodes is needed for the proposed attack, which is the main issue. Also, Reference [5] develops a stealth attack on battery energy storage systems where attackers first train an ANN model based on the stealth data to generate output similar to the battery energy storage system. Most articles consider only one type of cyberattack. However, there is a chance that attackers use two different types of attacks simultaneously. Reference [6] shows how a hybrid attack using an integrity attack and an availability attack simultaneously can cause severe damage to the system. Reference [7] studies a hybrid FDI and DOS attack on thermostatically controlled loads of in-building microgrids.

Only a few works have studied hybrid cyberattacks, and hybrid cyberattacks on the Volt-Var control system have not been sufficiently studied. Previous studies have used different defense methods for FDI and DOS attacks. However, machine learning methods work successfully for both FDI and DoS attacks. For FDI attack detection and mitigation, stochastic-based, ANN-based, state observer-based, statistical, and score-based solutions have been proposed. Reference [8] develops a mitigation method for PV Volt-Var control using an error-based local setting solution for set points of voltage regulation controller, and it uses a stochastic optimal solution for state estimation and bad data detection. Reference [9] proposed an artificial neural network ANN-based method for detecting and mitigating the FDI attack. An ANN estimates the output DC current in a DC microgrid to de-centrally detect and remove false data injection values for microgrids. Reference [10] develops a method using ANN for detecting and mitigating the FDI attack on DC microgrid inverter control signals. A system using ANN replaces falsified data in real-time with estimated data. Reference [11] uses a one-class detection model for detecting cyberattacks and a multilayer LSTM network for attack diagnosis for a PV farm. Reference [12] uses state observer in DC microgrid inverters to estimate and reconstruct the False data injection attack signal on measurements. The constructed signal is used to cancel the attack signal. This method can also be applied to inverter-based DERs to detect FDI attacks. Reference [13] uses data-driven methods based on raw electrical waveform data for detecting cyberattacks in grids with PVs. The score method with a statistical structure and no previous training is used to detect intrusion. Reference [14] simulates a scenario under FDI attacks that cause voltage disturbances in the grid with DERs and proposes a method using time stamps to detect and mitigate FDI. Reference [15] studies various scenarios of FDI attacks on Distributed renewable energy resources in a microgrid. Reference [16] studies FDI attacks on inverter-based resources in microgrids using PSCAD. Reference [17] cybersecurity assessment of solar PV units with reactive power capability.

Attackers continuously send packets, such as greeting packages, to fill the RTU's queue capacity and cause data loss in DoS attacks. The attack can be detected and mitigated by disconnecting any device using more than a certain percentage of the capacity of the queue of RTU. Moreover, ANN-based methods are used for DoS attack detection and mitigation. Reference [18] uses a multilayer data-driven cyberattack detection system using four classification models trained on network and host system data to detect DoS and Man-in-the-middle attacks. Reference [19] proposed two parallel support vector machines to detect DoS attacks. Reference [20] uses a deep convolutional neural network trained using real data to detect distributed DoS attacks over 5G projects. Reference [18] applies the finite-time observer technique to combine partial observers to modify finite-time partial observer-based controllers resilient to DoS attack.

This paper studies the hybrid attack, DoS, and FDI attacks simultaneously on a linearized Volt-Var control considering the unbalanced distribution voltage grid. The physical layer is implemented in PSCAD, the cyber layer is implemented using MATLAB Simulink, and the physical and the cyber layer are linked using the PSCAD cosimulation environment and Matrikon OPC for linking the cyber and physical layers. The proposed method successfully detects and mitigates two cases of hybrid cyberattacks. The main contribution of this paper is shown in the following:

• Hybrid attack effects are studied on the Volt-Var control system for the first time.

• For DoS attack detection and mitigation, an algorithm and an ANN-based method have been merged.

• The proposed method is tested for various cases to test the proposed method.

## Volt-Var control formulation

The Volt-Var optimization problem uses widely in the distribution grid to keep the voltage within the nominal range and reduce the loss of the distribution system. Since the voltage in the distribution grid is highly unbalanced, calculations must be done separately for all three phases. Reference [21] implemented a linearized Volt-Var optimization. The linearized Volt-Var optimization problem formulation is shown in the following:

$$\min J = \sum P_{ij}^{pp}(t) \, \forall i \in N \qquad (2)$$

where (2) is the objective function. The real and reactive power equality constraints are shown in the following:

$$P_{ij}^{pp}(t) = P_{jk}^{pp}(t) + P_{load,j}^{pp}(t) - \alpha_i^{DG} P_{DG,i}^{pp}(t) \quad \forall i \in N \qquad (3)$$

$$Q_{ij}^{pp}(t) = Q_{jk}^{pp}(t) + Q_{load,j}^{pp}(t) - \alpha_i^{DG} Q_{DG,i}^{pp}(t) - b_i^{CAP} Q_{CAP,i}^{pp}(t) \quad \forall i \in N \qquad (4)$$

$$b_i^{CAP} \in \{0,1\}, \alpha_i^{DG} \in \{0,1\} \qquad (5)$$

where, $P_{ij}^{pp}(t)$ is the real power flow of line between nodes $i$ and $j$ of phase $p$, $P_{load,j}^{pp}(t)$ is the load phases $p$ real power of node $j$, $P_{DG,i}^{pp}(t)$ is the distributed generated real power of phases $p$ at node $j$, $Q_{ij}^{pp}(t)$ is the reactive power flow of line between nodes $i$ and $j$ of phase $p$, $Q_{load,j}^{pp}(t)$ is the load phases $p$ reactive power of node $j$, $Q_{DG,i}^{pp}(t)$ is the distributed generation reactive power of phases $p$ at node $j$, and $Q_{CAP}^{pp}(t)$ is the capacitor bank injected reactive power at phase $p$ at node $j$. The capacitor reactive power equation is:

$$Q_{CAP,i}^{pp}(t) = u_{tap,i}^p(t) * Q_{CAP,i}^{rated,pp} v_i^p(t) \qquad (6)$$

where $v_i^p(t)$ represents voltage of phase $p$ at node $i$. The linearized Kirchhoff's Voltage Law (KVL) is:

$$v_j^p(t) = v_i^p(t) - \sum 2Re\left[S_{ij}^{pq}(t)\left(z_{ij}^{pq}\right)^*\right] \quad \forall j \in Y_i \tag{7}$$

where $v_j^p(t)$ represents voltage of phase $p$ at node $j$, $S_{ij}^{pq}(t)$ is the complex power of phases $p$ and $q$ of the line between nodes $i$ and $j$, and $z_{ij}^{pq}(t)$ is the lines impedance of phases $p$ and $q$ of the line between nodes $i$ and $j$. The OLTC equation is:

$$v_s^p(t) = \gamma_i^{OLTC,p} v_i^p(t) \tag{8}$$

$$\gamma_i^{OLTC,p} = \sum_{i=1}^{32} B_i u_{tap,i}^p(t), \sum_{i=1}^{32} u_{tap,i}^p(t) = 1 \tag{9}$$

where $u_{tap,i}^p(t)$ is phase p transformer tap positions. The limitations of DG output reactive power is:

$$-\sqrt{\left(S_{DG,i}^{rated,p}\right)^2 - \left(P_{DG,i}^p\right)^2} =< \alpha_i^{DG} Q_{DG,i}^p \leq \sqrt{\left(S_{DG,i}^{rated,p}\right)^2 - \left(P_{DG,i}^p\right)^2} \tag{10}$$

where $S_{DG,i}^{rated,p}$ is the rated DG power at phase p. The nodes voltage limitation equation is:

$$(V_{min})^2 \leq v_i^p(t) \leq (V_{max})^2 \quad \forall i \in N \tag{11}$$

where $V_{max}$ is the maximum voltage and $V_{min}$ is the minimum voltage.

## Proposed Method

A detection and mitigation method is proposed for the hybrid cyberattack on the Volt-Var control system, simultaneously applying DoS and FDI attacks. In an artificial neural network (ANN), information is processed through interconnected layers of neurons, where Each neuron receives input from other neurons, processes this input using a set of, and generates an output signal for other neurons. ANNs are one type of machine learning method, and previous studies show the effectiveness of ML-based methods for the detection and mitigation of cyberattacks [9]–[11], [13], [14]. Moreover, an ANN-based algorithm is implemented in the cyber layer of the control center to mitigate the hybrid attack.

DoS attack is detected if a user uses over 60% of the feeder remote terminal unit capacity. It is mitigated by replacing the lost value estimated by the ANN and disconnecting the attackers. The FDI is detected if the difference between the estimated value by ANN and the received measurement packet is more than 10% since the accuracy of the ANN is about 90.3%. If the difference between the estimated and the received value is greater than 10%, the estimated value is replaced with the received measurements. The proposed method replaces the falsified data packet before hitting the physical layer. The ANN is trained using a dataset consisting of a loading factor between 0.5 to 1.5 with 0.1 steps.

A modified IEEE 13-bus test feeder system is simulated in PSCAD shown in Fig. 1. A distributed generation is added to node 671 [21]. The cyber layer is implemented using D/D/m/K queuing and FIFO [22]. It is implemented in MATLAB Simulink and is shown in Fig. 2. OPC Matrikon transfers data bidirectionally among the power and cyber layer, and the relation between the cyber layer, physical layer, and OPC Matrikon is shown in Fig. 3. The efficiency of the proposed method is shown for two cases in the simulation results.

Paper number 1203
SC D2 – Information systems and telecommunication
Stream 2. Developing practices, functionalities and applications

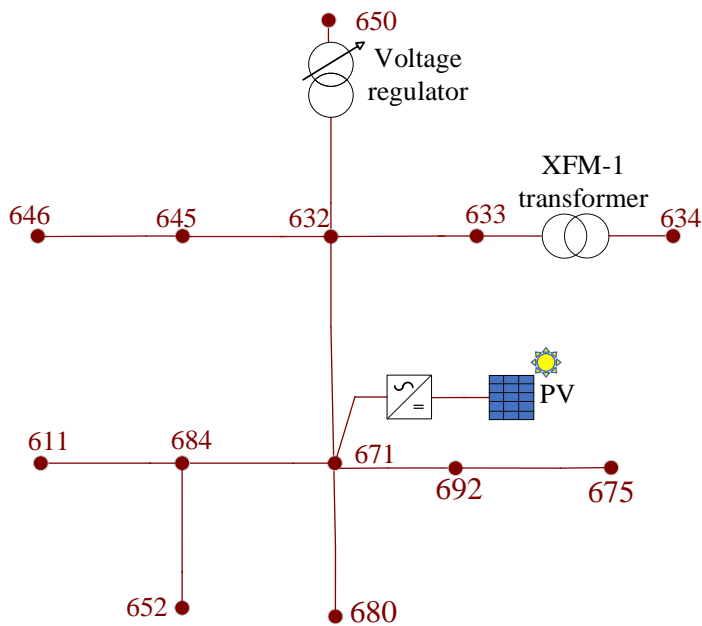

**Fig. 1.** The IEEE 13-bus test feeder system.

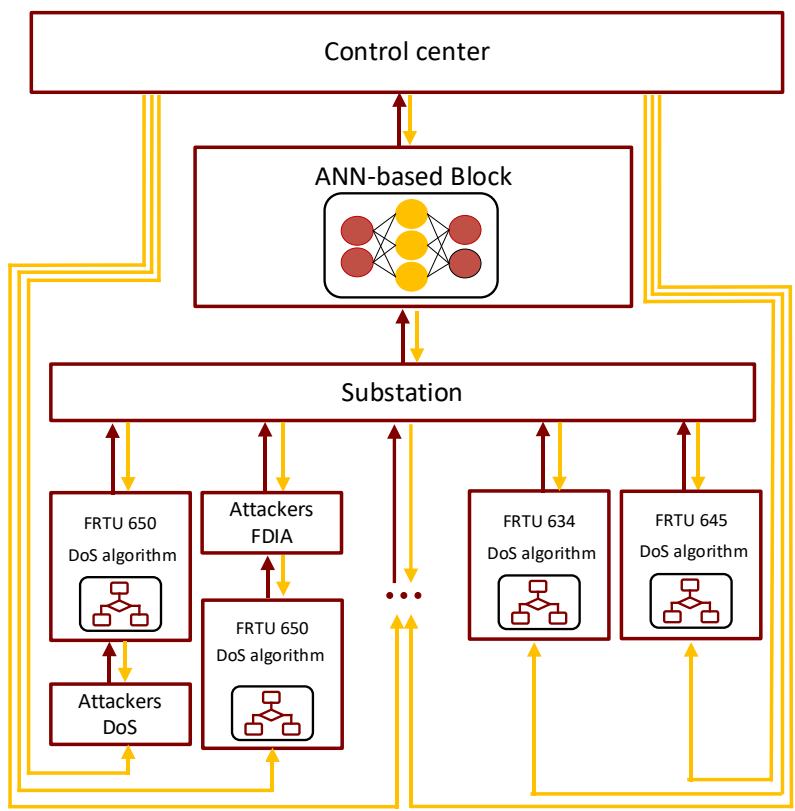

**Fig. 2.** The cyber layer models.

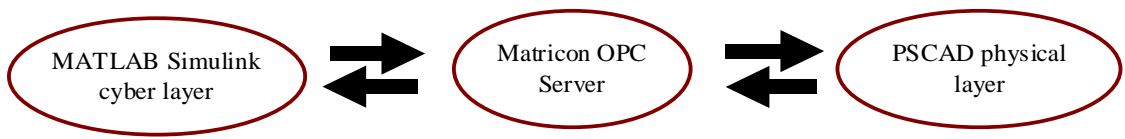

**Fig. 3.** The cyber and physical layer connection.

## Simulation Results

Two simulation studies are done on the modified IEEE 13-bus test feeder system to show the effectiveness of the proposed method. The voltage waveform for both cases, with and without mitigation, is shown for each case.

### First Scenario

The loading factor is 0.5 for all loads before the hybrid cyberattacks launch. FDIA is applied to the real and reactive power value of loads connected to nodes 680 and 671. The cyberattack changes the real and reactive value of the load connected to node 680 to 500 kW and 500 kVAr, respectively, for all three phases. The real and reactive part of the load connected to node 671 increases by 60%, and the DoS attack is applied to node 652. The hybrid attack is applied at t = 3 s. If the attack does not mitigate, the OLTC increases the voltage, and more reactive power is injected into the grid, causing an overvoltage. Fig. 4 shows node 671 voltages for a system without detection and mitigation, and Fig. 5 shows node 671 voltages for a system with the proposed detection and mitigation method. Fig. 4 shows the voltage of phase b reaches 1.11 pu and violates the limit range. However, it increases by 1% and does not violate the limit range in Fig. 5.

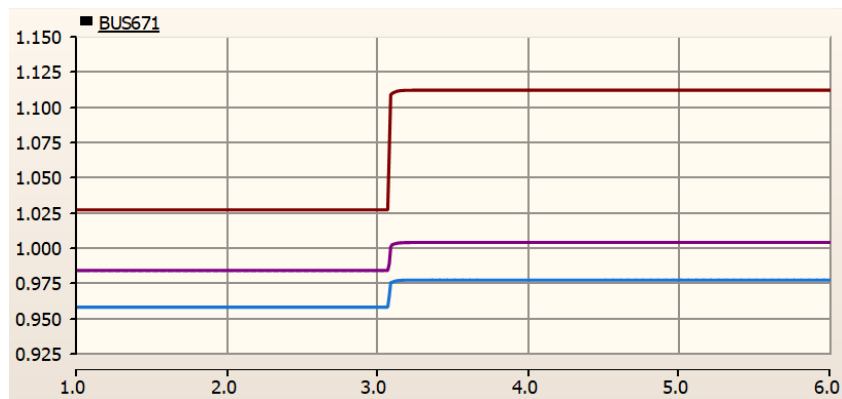

**Fig. 4.** Node 671 voltage without any detection and mitigation and hybrid attacks applied on nodes 680, 671, and 652 at t=3 s.

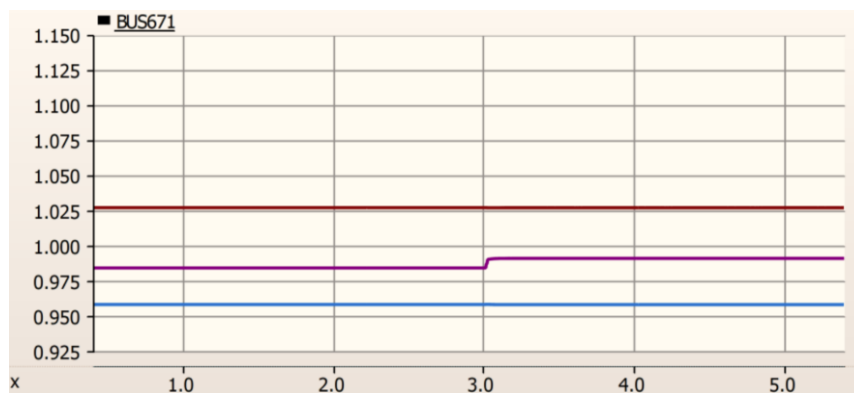

**Fig. 5.** Node 671 voltage with the proposed detection and mitigation methods and hybrid attacks applied on nodes 680, 671, and 652 at t=3 s.

### Second Scenario

The loading factor is 0.5 for all loads before the hybrid cyberattacks launch. FDIA is applied to the real and reactive power value of loads connected to nodes 680, 692, and 632. The DoS attack is applied to node 633. The cyberattack changes the real and reactive value of the load connected to phases a, b, and c of node 680 to 500 kW and 500 kVAr, respectively. It changes the real and reactive value of the load connected to phases a and c of node 692 to 100 kW and 100 kVAr, respectively. The hybrid attack is applied at t = 3 s. If the attack does not mitigate, the OLTC increases the voltage, and more reactive power is injected into the grid, causing an overvoltage. Fig. 6 shows node 671 voltages for a system without detection and mitigation, and Fig. 7 shows node 671 voltages for a system with the proposed detection and mitigation method. Fig. 6 shows the voltage of phase b reaches 1.11 pu and violates the limit range. However, it does not violate the range in Fig. 7.

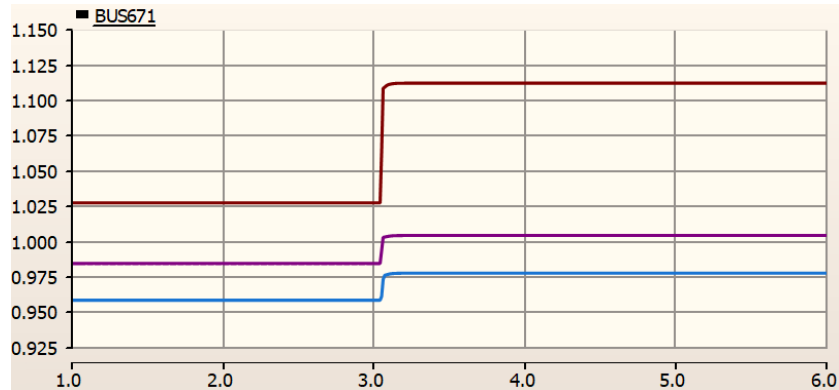

**Fig. 6** Node 671 voltage without any detection and mitigation and hybrid attacks applied on nodes 680, 692, and 632 at t=3 s.

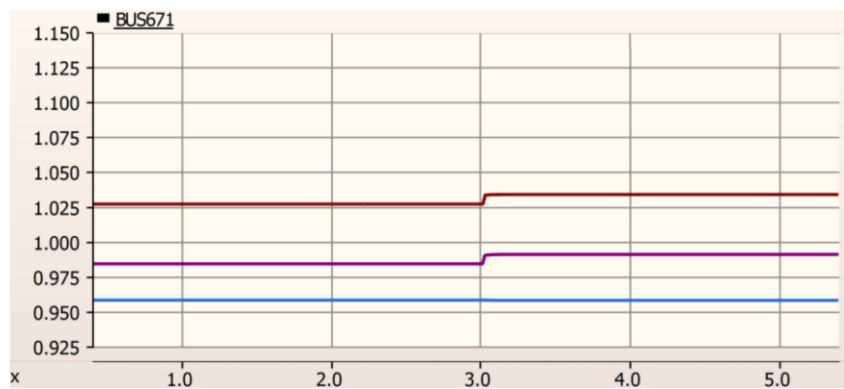

**Fig. 7.** Node 671 voltage with the proposed detection and mitigation methods and hybrid attacks applied on nodes 680, 692, and 632 at t=3 s.

For both cases, the algorithm mitigates the hybrid cyberattacks successfully.

### Reliability of the Detection Method

The training process of the ANN can affect the reliability and accuracy of the proposed method. The following are among the considerations that should be taken into account to evaluate the performance of the proposed method:

- Avoid overfitting the ANN: If the ANN is overfitted, it will have a high accuracy for the training data, but the accuracy will deteriorate for new data that the ANN has not used before.
- Expected number of compromised nodes: The distribution grid typically has thousands of nodes, and it is impractical for an attacker to compromise all these nodes. However, a higher number of compromised nodes can decrease the accuracy of the proposed method. In an example scenario, we studied how the possible voltage of each node changes as the number of compromised nodes increases. In this scenario, each compromised node's real power measurement is increased by 100 kW, and its reactive power measurement is increased by 100 kVAr. In each case, we studied different combinations of compromised nodes and recorded the maximum voltage. Fig. 8 shows the maximum voltages experienced vs. the number of compromised nodes. As shown, as the number of compromised nodes increases, the maximum voltage also increases. This adversely affects the reliability of the proposed method.

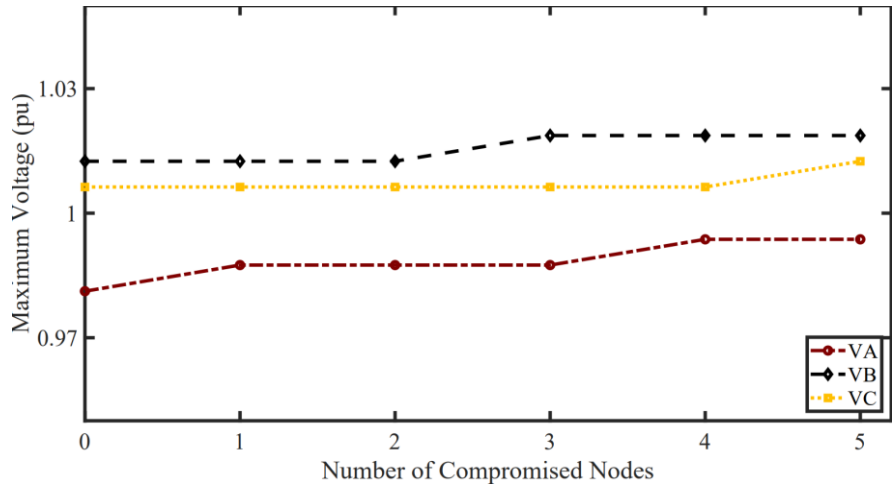

**Fig. 8.** Maximum voltage of each phase for the different number of compromised nodes.

## Conclusion

Cybersecurity is a major concern since attackers can use the internet-based infrastructure to cause disturbances and power outages. Attackers can use DoS and FDI attacks simultaneously, known as hybrid attacks, to cause more damage. Hybrid attacks can be applied to Volt-Var control resources and cause voltage disturbances. Proper detection and mitigation method are proposed for hybrid attacks on the Volt-Var control system using an ANN-based algorithm. A proper cyber-physical system is used in this work. The physical layer is implemented in PSCAD, the cyber layer is implemented using MATLAB Simulink, and the physical and the cyber layer are linked using the PSCAD cosimulation environment and Matrikon OPC for linking the cyber and physical layers. The proposed method successfully detects and mitigates two cases of hybrid cyberattacks as tested using the IEEE 13-bus test feeder system.

# Bibliography

[1] M. Mahmud, M. Hossain, and H. Pota, "Analysis of voltage rise effect on distribution network with distributed generation," IFAC Proceedings Volumes, vol. 44, no. 1, pp. 14 796–14 801, Jan. 2011, 18th IFAC World Congress.

[2] S. N. Salih and P. Chen, "On coordinated control of oltc and reactive power compensation for voltage regulation in distribution systems with wind power," IEEE Transactions on Power Systems, vol. 31, no. 5, pp. 4026–4035, Sept. 2016.

[3] Y. Isozaki, S. Yoshizawa, Y. Fujimoto, H. Ishii, I. Ono, T. Onoda, and Y. Hayashi, "Detection of cyber attacks against voltage control in distribution power grids with pvs," IEEE Transactions on Smart Grid, vol. 7, no. 4, pp. 1824–1835, Jul. 2016.

[4] D. Choeum and D.-H. Choi, "Oltc-induced false data injection attack on volt/var optimization in distribution systems," IEEE Access, vol. 7, pp. 34 508–34 520, Mar. 2019.

[5] M. Pasetti, P. Ferrari, P. Bellagente, E. Sisinni, A. O. de Sa, C. B. d. ´ Prado, R. P. David, and R. C. S. Machado, "Artificial neural networkbased stealth attack on battery energy storage systems," IEEE Transactions on Smart Grid, vol. 12, no. 6, pp. 5310–5321, Aug. 2021.

[6] H. Tu, Y. Xia, C. K. Tse, and X. Chen, "A hybrid cyber attack model for cyber-physical power systems," IEEE Access, vol. 8, pp. 114 876– 114 883, Jun. 2020.

[7] Y. Wan, C. Long, R. Deng, G. Wen, X. Yu, and T. Huang, "Distributed event-based control for thermostatically controlled loads under hybrid cyber attacks," IEEE Transactions on Cybernetics, vol. 51, no. 11, pp. 5314–5327, Mar. 2021.

[8] "Centralized volt–var optimization strategy considering malicious attack on distributed energy resources control," IEEE Transactions on Sustainable Energy, vol. 9, no. 1, pp. 148–156, Jun. 2018.

[9] M. R. Habibi, S. Sahoo, S. Rivera, T. Dragicevi ˇ c, and F. Blaabjerg, ´ "Decentralized coordinated cyberattack detection and mitigation strategy in dc microgrids based on artificial neural networks," IEEE Journal of Emerging and Selected Topics in Power Electronics, vol. 9, no. 4, pp. 4629–4638, Aug. 2021.

[10] M. R. Habibi, H. R. Baghaee, F. Blaabjerg, and T. Dragicevi ˇ c, "Secure ´ control of dc microgrids for instant detection and mitigation of cyberattacks based on artificial intelligence," IEEE Systems Journal, vol. 16, no. 2, pp. 2580–2591, Jun. 2022.

[11] F. Li, Q. Li, J. Zhang, J. Kou, J. Ye, W. Song, and H. A. Mantooth, "Detection and diagnosis of data integrity attacks in solar farms based on multilayer long short-term memory network," IEEE Transactions on Power Electronics, vol. 36, no. 3, pp. 2495–2498, Mar. 2021.

[12] A. Cecilia, S. Sahoo, T. Dragicevi ˇ c, R. Costa-Castell ´ o, and F. Blaabjerg, ´ "On addressing the security and stability issues due to false data injection attacks in dc microgrids—an adaptive observer approach," IEEE Transactions on Power Electronics, vol. 37, no. 3, pp. 2801–2814, Sep. 2022.

[13] F. Li, R. Xie, B. Yang, L. Guo, P. Ma, J. Shi, J. Ye, and W. Song, "Detection and identification of cyber and physical attacks on distribution power grids with pvs: An online high-dimensional data-driven approach," IEEE Journal of Emerging and Selected Topics in Power Electronics, vol. 10, no. 1, pp. 1282–1291, Sep. 2022.

[14] A. K. Jain, N. Sahani, and C.-C. Liu, "Detection of falsified commands on a der management system," IEEE Transactions on Smart Grid, vol. 13, no. 2, pp. 1322–1334, Dec. 2022.

[15] A. Mohammadhassani, A. Teymouri, A. Mehrizi-Sani, and K. Tehrani, "Performance Evaluation of an Inverter-Based Microgrid Under Cyberattacks," 2020 IEEE 15th International Conference of System of Systems Engineering (SoSE), Budapest, Hungary, 2020, pp. 211-216, doi: 10.1109/SoSE50414.2020.9130524.

[16] A. Mohammadhassani, Y. Akbar, A. Mehrizi–Sani and H. Wang, "Cyber Vulnerability Assessment of Microgrids with 5G- Enabled Distributed Control," 2022 IEEE Power & Energy Society General Meeting (PESGM), Denver, CO, USA, 2022, pp. 01-05, doi: 10.1109/PESGM48719.2022.9916971.

[17] A. Teymouri, A. Mehrizi-Sani and C. -C. Liu, "Cyber Security Risk Assessment of Solar PV Units with Reactive Power Capability," IECON 2018 - 44th Annual Conference of the IEEE Industrial Electronics Society, Washington, DC, USA, 2018, pp. 2872-2877, doi: 10.1109/IECON.2018.8591583.

[18] F. Zhang, H. A. D. E. Kodituwakku, J. W. Hines, and J. Coble, "Multilayer data-driven cyberattack detection system for industrial control systems based on network, system, and process data," IEEE Transactions on Industrial Informatics, vol. 15, no. 7, pp. 4362–4369, Jul. 2019.

[19] J. Qian, X. Du, B. Chen, B. Qu, K. Zeng, and J. Liu, "Cyber-physical integrated intrusion detection scheme in scada system of process manufacturing industry," IEEE Access, vol. 8, pp. 147 471–147 481, Aug. 2020.

[20] B. Hussain, Q. Du, B. Sun, and Z. Han, "Deep learning-based ddosattack detection for cyber–physical system over 5g network," IEEE Transactions on Industrial Informatics, vol. 17, no. 2, pp. 860–870, Feb. 2021.

[21] R. R. Jha, A. Dubey, C.-C. Liu, and K. P. Schneider, "Bi-level volt-var optimization to coordinate smart inverters with voltage control devices," IEEE Transactions on Power Systems, vol. 34, no. 3, pp. 1801–1813, Jan. 2019.

[22] A. Stefanov, C.-C. Liu, M. Govindarasu, and S.-S. Wu, "Scada modeling for performance and vulnerability assessment of integrated cyber–physical systems," International Transactions on Electrical Energy Systems, vol. 25, no. 3, pp. 498–519, Mar. 2015